\documentclass[aps,prb,reprint]{revtex4-1}%
\usepackage{epsfig,pslatex,latexsym,times}
\usepackage{bm, float, lipsum}
\usepackage{amsmath}
\usepackage{amsfonts}
\usepackage{amssymb}
\usepackage{mathrsfs}
\usepackage{color}
\usepackage{graphicx}%
\providecommand{\U}[1]{\protect\rule{.1in}{.1in}}

\begin{document}
\author{N. J. Harmon}
\email{harmon.nicholas@gmail.com} 
\affiliation{Department of Physics and Astronomy, University of Iowa, Iowa City, Iowa
52242, USA}
\affiliation{Department of Physics, University of Evansville, Evansville, Indiana
47722, USA}
\author{M. E. Flatt\'e}
\email{michael\_flatte@mailaps.org} 
\affiliation{Department of Physics and Astronomy, University of Iowa, Iowa City, Iowa
52242, USA}
\affiliation{Department of Applied Physics, Eindhoven University of Technology, Eindhoven 5612 AZ, The Netherlands}
\date{\today}
\title{Theory of  oblique-field magnetoresistance from spin centers in three-terminal spintronic devices}
\begin{abstract}
We present a general stochastic Liouville theory of electrical transport across a barrier between two %magnetic 
conductors that occurs via sequential hopping through a single defect's spin-$0$ to spin-$1/2$ transition.% level energetically within the bias window. 
%When one magnetic and one non-magnetic conductor are used, under the bias conditions for spin extraction, 
We find magnetoconductances 
similar to Hanle features (pseudo-Hanle features) that   originate from Pauli blocking without spin accumulation, and also predict that 
%Under bias during spin injection we predict that 
 evolution of the defect's spin modifies the conventional Hanle response, producing  %from that expected without spin centers, and that 
 an inverted Hanle signal from spin center evolution. 
%modified Hanle response due to spin evolution on the defect. Also a inverted Hanle signal is derived from the spin injection.
We propose studies in oblique magnetic fields that would %an experimental procedure, in which the external magnetic field angle is varied, which 
unambiguously determine if %an observed field-dependent voltage or current 
a magnetoconductance results from  spin-center assisted transport.%, and reinterpret recent oblique field magnetoresistance data.% from He \emph{et al} J. Appl. Phys. {\bf 119}, 113902 (2016).
%The described magnetic-field effects require charge current, and so are not expected to obscure results from non-local measurements.
\end{abstract}
\maketitle	

% \begin{itemize}\item \href{https://orcid.org/0000-0000-0000-0000}{\textcolor{orcidlogocol}{\aiOrcid} \hspace{2mm} orcid.org/0000-0000-0000-0000}\end{itemize}

\emph{Introduction.}---  %Three-terminal (3T) spin injection and spin extraction experiments continue to perplex the spintronics community \cite{Txoperena2016}. 
A current flowing through a magnetic conductor, with carriers flowing into (spin injection) or out of (spin extraction) a nonmagnetic conductor, can produce a nonequilibrium spin population in the nonmagnetic material\cite{Johnson1985} that can be sensed through a spin-sensitive voltage. An applied magnetic field, intended to precess these  nonequilibrium spins and thereby reduce the voltage (the Hanle effect or HE\cite{Johnson1988}), often produces surprising results challenging to interpret using reasonable spin coherence times\cite{Tran2009, Jansen2012a, Txoperena2013, Inoue2015,Txoperena2016}. Anomalous Hanle features (the inverted Hanle effect or IHE) were found with parallel applied field and magnetization; conventional interpretations attribute IHE to magnetic fringe fields from a non-uniform interface, however detailed structural measurements of the interface failed to correlate these features with measured non-uniformity. 
For currents and voltages measured using the same contact (three-terminal, or 3T measurements), features mimicking the HE and IHE were found\cite{Txoperena2014, Inoue2015}, originating from magnetic-field-dependent transport through spin centers. However, agreement between some 3T and four-terminal (4T) non-local experiments indicates that the impurity effect does not always dominate over direct tunneling \cite{Lou2006, Lou2007, Aoki2012}.
%Discriminating between direct and sequential tunneling will assist in understanding and interpreting 3T signatures.
%The field-angle dependence of magnetoresistance may distinguish between spin transport involving spin centers,% spin injection/extraction and spin-center sequential tunneling, 
%since the mechanisms behind these two transport phenomena are different.\red{$\leftarrow$ This sentence was confusing to me; is there a word missing?}%, however the first analysis along these lines did not identify the key features of the measurement which would differ, and hence could not distinguish betweeen the two models\cite{He2016}.  %\textcolor{purple}{Yes.  He2016 only analyzed the high-field current and not the relative change in current between hi and 0 field. Their fits to the high-current data with both the SD and spin injection models was unable to distinguish which model was better.}

Here we analyze sequential spin-center-mediated tunneling between different magnetic or non-magnetic leads in both the spin injection and spin extraction regimes using the stochastic Liouville equation (SLE) formalism.
The SLE framework is highly adaptable to many physical systems, including spin-oriented tunneling through spin centers\cite{Harmon2018}. We apply the formalism to sequential tunneling involving a spin center  that  alternates between spin zero and spin one-half as its charge state changes.
In the spin extraction regime, our formalism confirms previous studies that demonstrated supposed HE and IHE to be the result of a Pauli blockade \cite{Song2014, Inoue2015}, which we now refer to as ``pseudo HE" and ``pseudo IHE'' phenomena. To indicate the experimental geometry without assigning a mechanism we refer to ``Hanle measurements'' or ``inverted Hanle measurements''.
%We go further though by investigating the spin center's role in spin injection. 
The combined effects of spin accumulation at the spin center and the coherent evolution of that spin accumulation lead to an altered spin injection process; we predict a previously undescribed, broad IHE (with same physical origin as the conventional HE) accompanying the known conventional HE (that can be broad or narrow). We will refer to these as ``conventional HE'' and ``conventional IHE'', distinct from the pseudo HE and pseudo IHE.
We analyze the oblique field data of Ref.~\onlinecite{He2016} to show that a %carefully defined
ratio of the angular dependent responses could, in principle, provide incontrovertible evidence of the impurity model by definitively ruling out the fringe-field mechanism for the inverted signal. 
%We cannot unambiguously specify where the energy of the relevant defect's charge occupancy transition, in Ref.~\onlinecite{He2016}, occurs within the energy range accessible by the voltage bias,
%but o
Our calculations predict  distinct magnetic field widths in the magneto-response that motivate further measurements to resolve the origin of the observed features.% questions. % \red{is there qualitative specificity?}
 Lastly we show that the effects described herein \emph{require charge current} and are thus not present for non-local (4T) measurements (or spin pumping). %This conclusion extends to spin pumping experiments; the spin centers, within the assumptions of our model, will not alter a spin current without an electrical bias.

\begin{figure}[ptbh]
 \begin{centering}
        \includegraphics[scale = 0.36,trim = 0 0 0 0, angle = -0,clip]{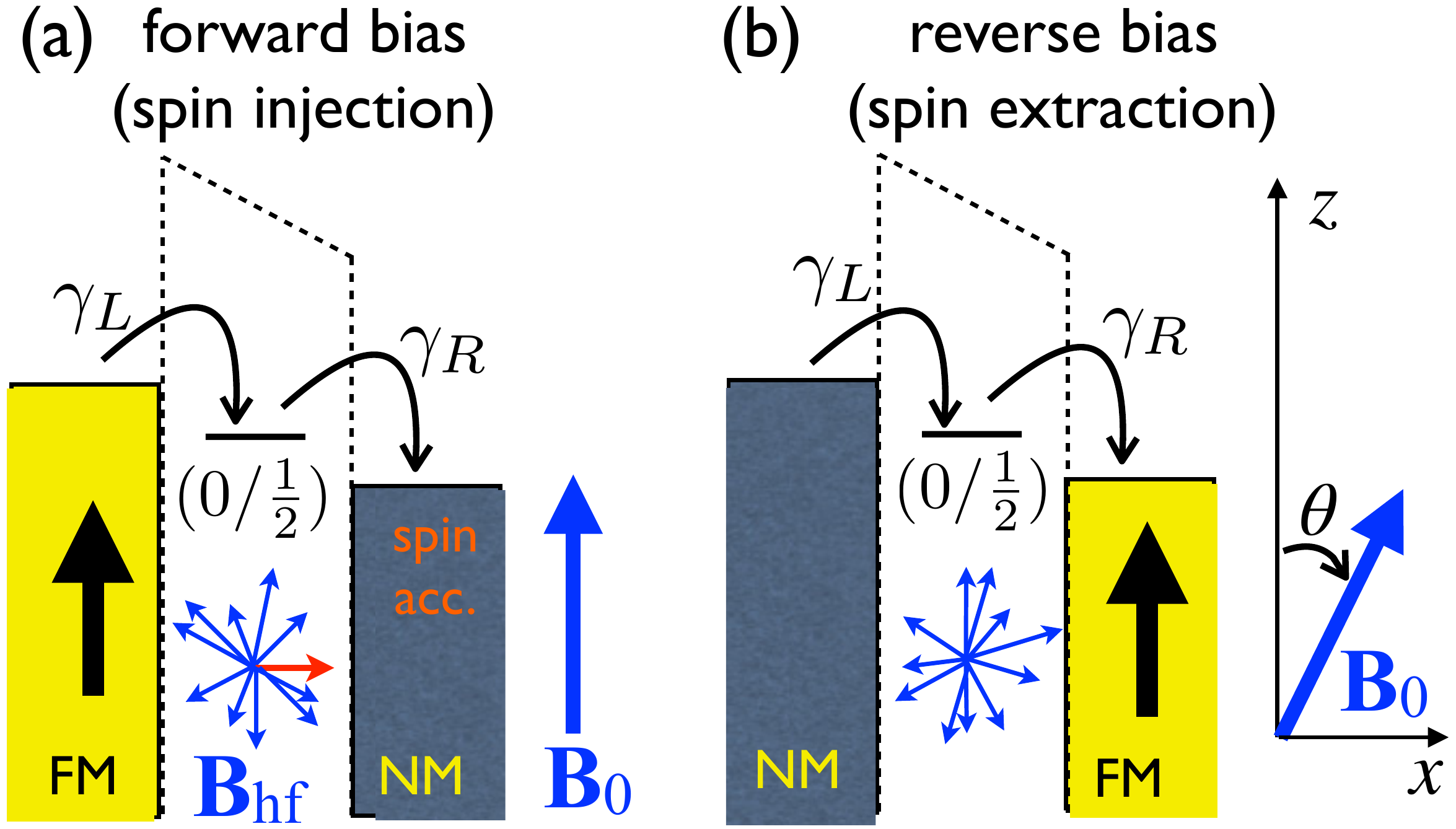}
        \caption[]
{Spin center-mediated transport between a ferromagnet (FM) and nonmagnetic metal (NM) for (a) spin injection and (b) spin extraction with interfacial spin centers possessing a transition level (0/$\frac{1}{2}$). The barrier trap in (a) is a spin transport center.  The barrier trap in (b) is a spin bottleneck center.
%at t located at the interface of the FM and NM. 
The IHE emerges from (a). %The applied field, $B_0$, is oriented parallel to the FM's magnetization ($\theta = 0$) so the conventional expectation is that there is no HE.
If the spin center possesses a nuclear spin moment, assumed to be randomly oriented, some portion (red arrow in (a)) of the hyperfine field will be perpendicular to $\bm{B}_0$. This $\perp \bm{B}_0$ component reduces the spin current; 
increasing $B_0$ \emph{reduces} the relative importance of $B_{hf}$ which leads to increasing spin accumulation in the NM. (b)  The applied magnetic field makes an angle $\theta$ with respect to the $z$-axis as shown in (b).  }\label{fig:impurityFig1}
        \end{centering}

\end{figure}

\emph{Interfacial Spin Centers.}---  This work treats spin injection and spin extraction between a non-magnetic material (NM) and a ferromagnet (FM); the charge and/or spin current passes through an interfacial barrier trap state through sequential hopping from one lead onto the trap, and then from the trap to the other lead. %. In each case we examine the role of sequential hopping through spin centers in the insulator. 
Figure~\ref{fig:impurityFig1}(a) shows spin injection under a forward bias.
%The barrier trap within the insulating barrier contains a transition level within the bias window that alternates between $s = 1/2$ and $s = 0$ as charge hops from left to right. 
The transition level of the spin center within the insulating barrier is labeled as (0/$\frac{1}{2}$) to denote the two possible spin states: $0$ here refers to the spin when no charge has hopped onto the level whereas $\frac{1}{2}$ is the spin when a charged has hopped onto the level.
Figure~\ref{fig:impurityFig1}(b) shows the same center in the spin extraction regime (reverse bias, or with the non-magnetic and ferromagnetic contacts reversed).  We differentiate between the two configurations of Fig.~\ref{fig:impurityFig1} by defining (a) a \emph{spin transport center} (since spin will accumulate in the NM) and (b) a \emph{spin bottleneck center} since a bottleneck forms at the right junction. For both situations, there are only the two designated spin states of the defect, differing by single charge carrier occupancy. The supplement discusses  barrier traps with a transition level ($\frac{1}{2}$/0), which under forward (reverse) bias yield charge and spin dynamics identical to the (0/$\frac{1}{2}$) level under reverse (forward) bias.\cite{supp} Thus the spin and charge currents under forward bias, when both  (0/$\frac{1}{2}$) and ($\frac{1}{2}$/0) centers are present, can be accurately described by combinations of the two configurations shown in Fig. \ref{fig:impurityFig1}, with $N_t$ spin transport centers and $N_b$ spin bottleneck centers. 
The total current %through $N_t$ ($N_b$) number of equivalent parallel $(0/\frac{1}{2})$ ($(\frac{1}{2}/0)$) transition levels
then is the sum of such currents:
\begin{eqnarray}\label{eq:current0}
i_{defect} &=& N_t i_t(\bm{P}_L = \bm{P}, \bm{P}_R = 0) + N_b i_b (\bm{P}_L = \bm{P}, \bm{P}_R =0),\nonumber \\ 
%\sum_j^{N_t} i_{t,j} + \sum_j^{N_b} i_{b,j} \\ \nonumber
&=& N_t i_t(\bm{P}_L = \bm{P}, \bm{P}_R = 0) + N_b i_t (\bm{P}_L = 0, \bm{P}_R =\bm{P}),
\end{eqnarray}
where $i_{defect} = i_{tot}-i_{dc} $ is the difference between the total current $i_{tot}$ and the direct (unaffected by the center) tunneling current  $i_{dc}$. The equivalency  $ i_b(\bm{P}_L =\bm{P}, \bm{P}_R = 0)\equiv i_t(\bm{P}_L = 0,\bm{P}_R =\bm{P})$ is derived described in the Supplement \cite{supp}. 
We concern ourselves only with $i_{defect}$ so our goal is to calculate $i_t(\bm{P}_L = \bm{P}, \bm{P}_R = 0)$ and $i_t(\bm{P}_L = 0,\bm{P}_R =\bm{P})$.

\emph{Theory.}--- Operators for  the static magnetizations of each electrode are $\hat{\bold{M}}_{L, R} = \frac{1}{2}(I + \bold{P}_{L, R}\cdot \bold{\sigma})$.
The time-evolution of the density matrix of the spin center, $\rho(t)$, is determined by  the SLE.
A 2$\times$2 matrix for the current $\hat{i}$ fully describes the flow of charge and coherent spin. Diagonal elements represent the movement of charge with  up or down spins, and off-diagonal elements describe the flow of charge with up and down spin superpositions. 
Charge currents are then $i=\text{Tr}\hat{i}$, and spin currents are $\bm{i}_{s,L(R)}=\text{Tr}\hat{i}_{L(R)}\bm{\sigma}$. 

The current operators are formulated by considering {\it e.g.}, in 
Fig.~\ref{fig:impurityFig1}(b), the probability for a charge to pass from the barrier trap to the FM to be $\frac{1}{2} ( 1 + \bm{P}_d(t) \cdot \bm{P}_R)$. The generalization of this probability for our current operator is $\hat{\bold{M}}_{R} \rho(t)$ which yields an expression for the `right' current (trap to right lead $R$)\cite{Harmon2018}
 \begin{equation}\label{eq:iR}
 \hat{i}_R(t) = \frac{e}{2} \gamma_R \Big[\hat{\bold{M}}_{R} \rho(t) + \rho(t)^{\dagger} \hat{\bold{M}}_{R}^{\dagger} \Big],
 \end{equation}
where the second term ensures hermiticity. 
 The `left' current (left lead $L$ to trap) can be derived in a similar fashion after constraining the center to be at most singly occupied:
 \begin{equation}\label{eq:iL}
\hat{i}_L(t) = e\gamma_L[1-\text{Tr}\rho(t)]\hat{\bold{M}}_{L}.
\end{equation}
Charge conservation demands that the `left' charge current equal the `right' charge current. 
%No such stipulation exists for the spin current, as the spin can evolve while on the defect site, which from a spin current perspective appears as a spin current source or sink.

To determine charge currents, spin currents, and spin accumulations,  the spin center's steady-state density matrix must be found. The center undergoes several interwoven processes (\emph{e.g.} charge hopping on and off, applied and local fields, charge and spin blocking) so unraveling the physics is not intuitive, however the SLE is well suited for  this type of problem. 
It avoids complications due to the choice of a spin quantization axis (see concluding remarks).
The  SLE \cite{Haberkorn1976}:
%\begin{widetext}
\begin{equation}\label{eq:SLE}
\frac{\partial\rho(t)}{\partial t} = \underbrace{-\frac{i}{{\hbar}} [\mathscr{H} , \rho(t)]}_{\text{coherent evolution}} -  \underbrace{\gamma_R \{ \hat{\bold{M}}_{R}, \rho(t)\}}_{\text{spin selection}}
 +  \underbrace{2\gamma_L [1 - \text{Tr}\rho(t)] \hat{\bold{M}}_{L}}_{\text{generation}} ,
 \end{equation}
%\end{widetext}
 where $\gamma_{R,L}$ are the spin dependent hopping rates to the right (left) electrode. 
 The spin Hamiltonian at the spin center site is $\mathscr{H} = {\frac{\hbar}{2} (\bm{b}_0 + \bm{b}_{hf})} \cdot \bm{\sigma} = \frac{1}{2} \bm{b} \cdot \bm{\sigma}$~ {where { $\bm{b}_0 = g \mu_B \bm{B}_0/\hbar$, $\bm{b}_{hf} = g \mu_B \bm{B}_{hf}/\hbar$,  $\bm{b} = \bm{b}_0 + \bm{b}_{hf}$, $\bm{B}_0$ is a uniform magnetic field, and $\bm{B}_{hf}$ is the hyperfine field at the spin center}. The hyperfine fields are assumed to be distributed as a gaussian function with width $\overline{b}_{hf}$.}
 The first term of the SLE represents the coherent evolution of the center's spin spin, the second term (curly braces are anti-commutators) denotes the spin-selective nature of tunneling into the FM. The third term describes hopping onto the impurity site from the left contact. We assume the spin lifetimes and coherence times of the spin ($T_1$ and $T_2$) are longer than the transport processes producing the current. %Typically $T_1$ times are extremely long for localized spins. Widely spaced spins do not interact with each other strongly, so we assume the $T_2$ is long as well, ignoring  spin relaxation and decoherence  of the trap spin.

\emph{Spin Extraction.}---  We now apply this theory to spin extraction [Fig.~\ref{fig:impurityFig1}(b)\cite{supp}].
The current,
\begin{equation}\label{eq:current}
i \equiv i(\bm{P}_R, 0) = e \frac{(1 - P_R^2 \chi(\bm{b}) ) \gamma_L \gamma_R}{(1 - P_R^2 \chi(\bm{b}) )\gamma_R+ 2 \gamma_L},
\end{equation}
with $\chi(\bm{b}) = (\gamma_R^2  + (\bm{b}\cdot \hat{P}_R)^2)/(\gamma_R^2 + b^2)$
% \begin{equation}
%\chi(\bm{B}) = \frac{\gamma_R^2  + (\bm{B}\cdot \hat{P}_R)^2}{\gamma_R^2 + B^2}.
% \end{equation}
 is  the same result found in Refs. \onlinecite{Song2014, Inoue2015}, but differs from {Ref.~\onlinecite{Yue2015b} (see concluding remarks). }

The spin polarization of the impurities, $\bm{P}_d = \text{Tr}(\bm{\sigma} \rho)$, is
 \begin{equation}
\bm{P}_d = - \frac{2\gamma_L}{\gamma_R (1 - P_R^2 \chi(\bm{b})) + 2 \gamma_L} \frac{\gamma_R^2 \bm{P}_R + \gamma_R \bm{b} \times \bm{P}_R + (\bm{b} \cdot \bm{P}_R) \bm{B}}{\gamma_R^2 + b^2}.
 \end{equation}
Spins parallel to the FM preferentially leave the barrier trap, thus a polarization opposite to the FM develops on the impurity site. This can be seen for $\bm{b}_{hf} = 0$ and $\bm{b}_0 || \bm{P}_R$, which leads to $\bm{P}_d \propto - \bm{P}_R$. This accumulation of defect spins, due to requiring spins leaving the defect to be parallel to $\bm{P}_R$, is called \emph{spin filtering} \cite{Harmon2018}.
There can be no spin accumulation in the NM since the impurity is only filled from the NM when it is empty; therefore no preferred spin is taken out of the NM. \cite{supp}

\emph{Spin Injection.}--- There is no spin filtering effect for the spin injection geometry [Fig.~\ref{fig:impurityFig1}(a)], so the current, $ i(\bm{P}_L, \bm{P}_R = 0) = e\gamma_L \gamma_R/(\gamma_R + 2 \gamma_L)$ is independent of $\bm{b}_{0} $.
The steady state impurity spin polarization,
 \begin{equation}\label{eq:defectspin}
\bm{P}_d = \frac{2\gamma_L}{\gamma_R + 2 \gamma_L} \left[\frac{\gamma_R^2 \bm{P}_L + \gamma_R \bm{b} \times \bm{P}_L + (\bm{b} \cdot \bm{P}_L) \bm{b}}{\gamma_R^2 + b^2}\right].
 \end{equation}
The spin current, $\bm{i}_s = e \gamma_R \bm{P}_d/2$, leads to an excess (above an assumed unpolarized background) spin polarization of NM carriers per unit volume, which we call the spin accumulation density, $\bm{p}_0$, at the interface with area $A$. The spin accumulation density decays further within the NM as $\bm{p} = \bm{p}_0 \exp(-x/\lambda_s)$ where $\lambda_s$ is the NM spin diffusion length. By setting the spin gained due to the spin current equal to the spin loss/evolution in NM,\cite{supp} $ \bm{i}_s/e = A \lambda_s  (\frac{1}{\tau_s} \bm{p}_0 - \bm{b}_0 \times \bm{p}_0)$, the spin accumulation {density} can be solved analytically for a general spin current:
\begin{equation}\label{eq:spinaccumulation}
\bm{p}_0 = \frac{\tau_s}{A  \lambda_s} \frac{\gamma_R}{2} \frac{\bm{P}_d + \tau_s \bm{b}_0 \times \bm{P}_d + \tau_s^2 (\bm{b}_0 \cdot \bm{P}_d) \bm{b}_0}{1 + \tau_s^2 b_0^2},
\end{equation}
 where the field dependence is hidden within $\bm{P}_d$ [see Eq.~(\ref{eq:defectspin}]. 
This expression is identical to expressions for the oblique Hanle effect for direct tunneling, replacing the spin polarization injected into the NM with the defect polarization $\bm{P}_d$.\cite{Meier1984}

Canting  the direction of the spin current away from the magnetization direction produces a previously unknown effective IHE.
The origin of the canting is seen by investigating Eq.~(\ref{eq:defectspin}) for  scenarios where $\bm{b}_{hf} || \bm{P}_L || \bm{b}_0 $ and $\bm{b}_{hf} \perp \bm{P}_L || \bm{b}_0 $.
When $\bm{b}_{hf} || \bm{P}_L || \bm{b}_0 $, $\bm{P}_d || \bm{P}_L$ (\emph{i.e.} no canting) and upon substituting Eq.~(\ref{eq:defectspin})  into Eq.~(\ref{eq:spinaccumulation}) we find no field dependence.
For $\bm{b}_{hf} \perp \bm{P}_L || \bm{b}_0 $, $\bm{P}_d$ develops components transverse to $\bm{P}_L$ which indicate canting and the component of $\bm{p} || \bm{P}_L$ will be smaller than for a parallel hyperfine field. As the applied field is increased the transverse hyperfine component's effect is minimized and the maximum $p_z$ is restored.
This is an IHE but in reality its origin is the same as the conventional HE except that a ``hidden" transverse field (a component of the hyperfine field $\perp \bm{P}_L || \bm{b}_0 $ ) exists at the trap site. This transverse hyperfine component rotates the spin that is injected into the NM which results in a smaller accumulation of $p_z$. Prior theories of inverted Hanle measurements have either been (1) solely attributed to the spatially inhomogeneous fringe fields of the ferromagnetic contact\cite{Dash2011}, 
or (2) assigned to the pseudo IHE\cite{Song2014,Txoperena2014,Inoue2015}. 

\begin{figure}[ptbh]
 \begin{centering}
        \includegraphics[scale = 0.5,trim = 0 0 0 0, angle = -0,clip]{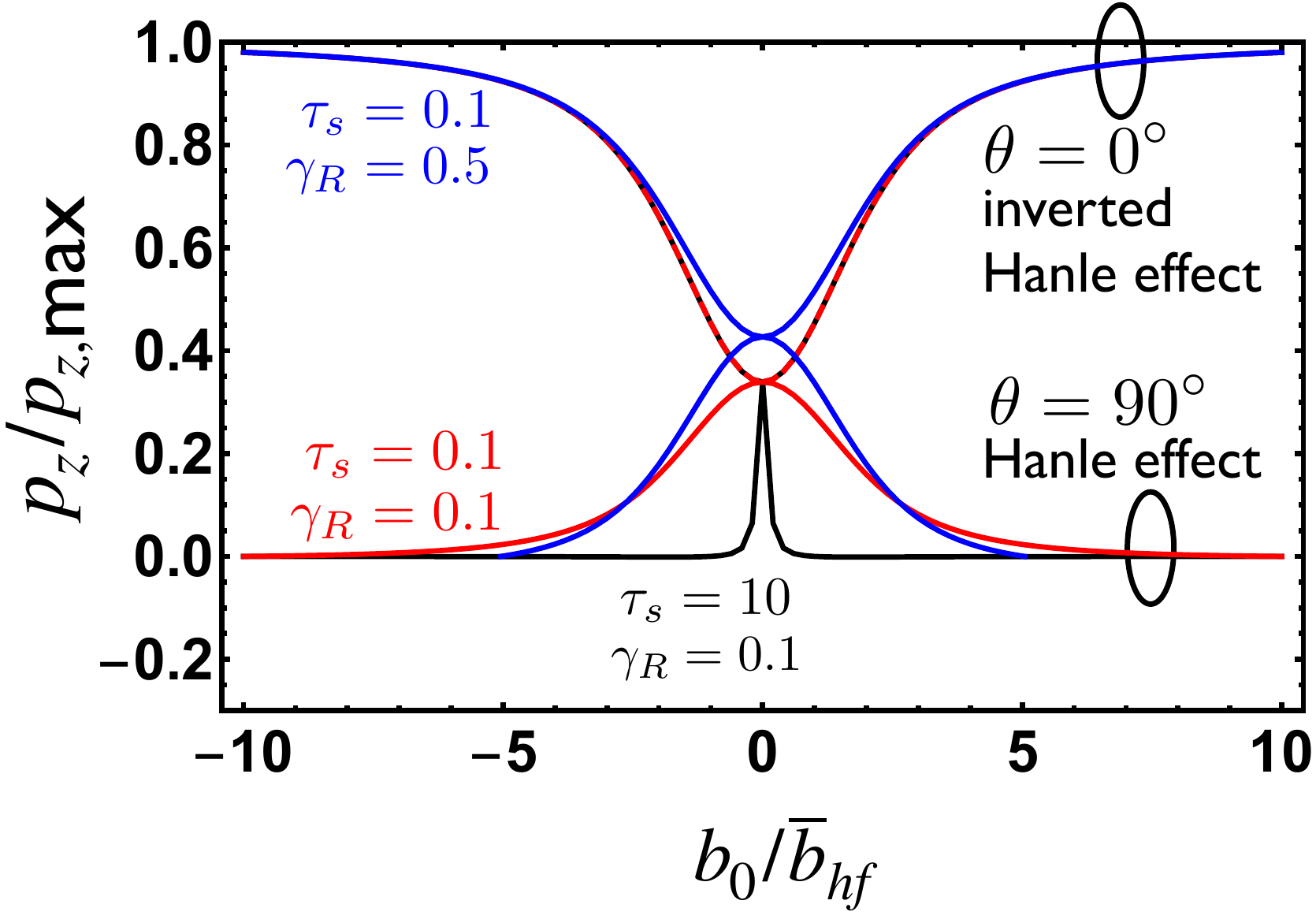}
        \caption[]
{Spin accumulation density parallel to $\hat{z}$ in NM, under conditions of spin injection [see Fig.~\ref{fig:impurityFig1}(a)], normalized to the maximum possible spin accumulation density, $p_{z,\text{max}} = \gamma_L \gamma_R \tau_s/A \lambda_s (\gamma_R + 2 \gamma_L)$. The conventional Hanle (inverted Hanle) curves are concave down (up) and  have widths sensitive (insensitive) to the spin relaxation time of traps, $\tau_s$. Colors (red, blue, and black) denote different  $\tau_s$'s and $\gamma_R$'s chosen in the calculation. The red curves are dashed to demonstrate that the red and black parameters yield negligibly different inverted Hanle curves.  $\gamma_L = 10$ and all rates are in units of $\overline{b}_{hf}$.}\label{fig:hanle}
        \end{centering}
\end{figure}

The hyperfine-field averaged NM spin accumulation is shown in Figure \ref{fig:hanle}. The combined dephasing evident in Eq.~(\ref{eq:defectspin}) and Eq.~(\ref{eq:spinaccumulation}) changes the conventional quadratic fall off in field to $p_z \sim 1/(\gamma_R^2 + b^2)(1 + \tau_s^2 b_0^2)$.
The width of the effects are primarily determined by  $\gamma_R$, $\overline{b}_{hf}$, or $\tau_s^{-1}$;
the width of the IHE is governed by the hyperfine field, and the HE by $\tau_s^{-1}$ for {$\tau_s^{-1}\ll \overline{b}_{hf}$}, leading to very different widths.
The discrepancy  should offer a means of distinguishing if this trap-mediated spin accumulation leads to observed spin voltages. However, if
{$\tau_s^{-1} \gg \overline{b}_{hf}$}, the widths are the same.

Experiments\cite{Aoki2012} on Fe/MgO/Si do observe a sharper peak of the HE superimposed upon a broader peak, as well as a broader inverted Hanle peak.
The spin lifetime of the narrow peak is  comparable to the one obtained in an accompanying nonlocal four-terminal experiment, which suggests an origin in direct tunneling.
The presence of similarly broad peaks in the Hanle and inverted Hanle curves point to additional charge hopping through spin bottleneck sites,  though contributions from hopping through spin transport sites  cannot be ruled out. Other experiments\cite{Sato2015} on Fe/SiO$_{2}$/Si  displayed  behavior in oblique fields which  suggests the importance of stray fields.

\begin{figure}[ptbh]
 \begin{centering}
        \includegraphics[scale = 0.25,trim = 0 0 0 0, angle = -0,clip]{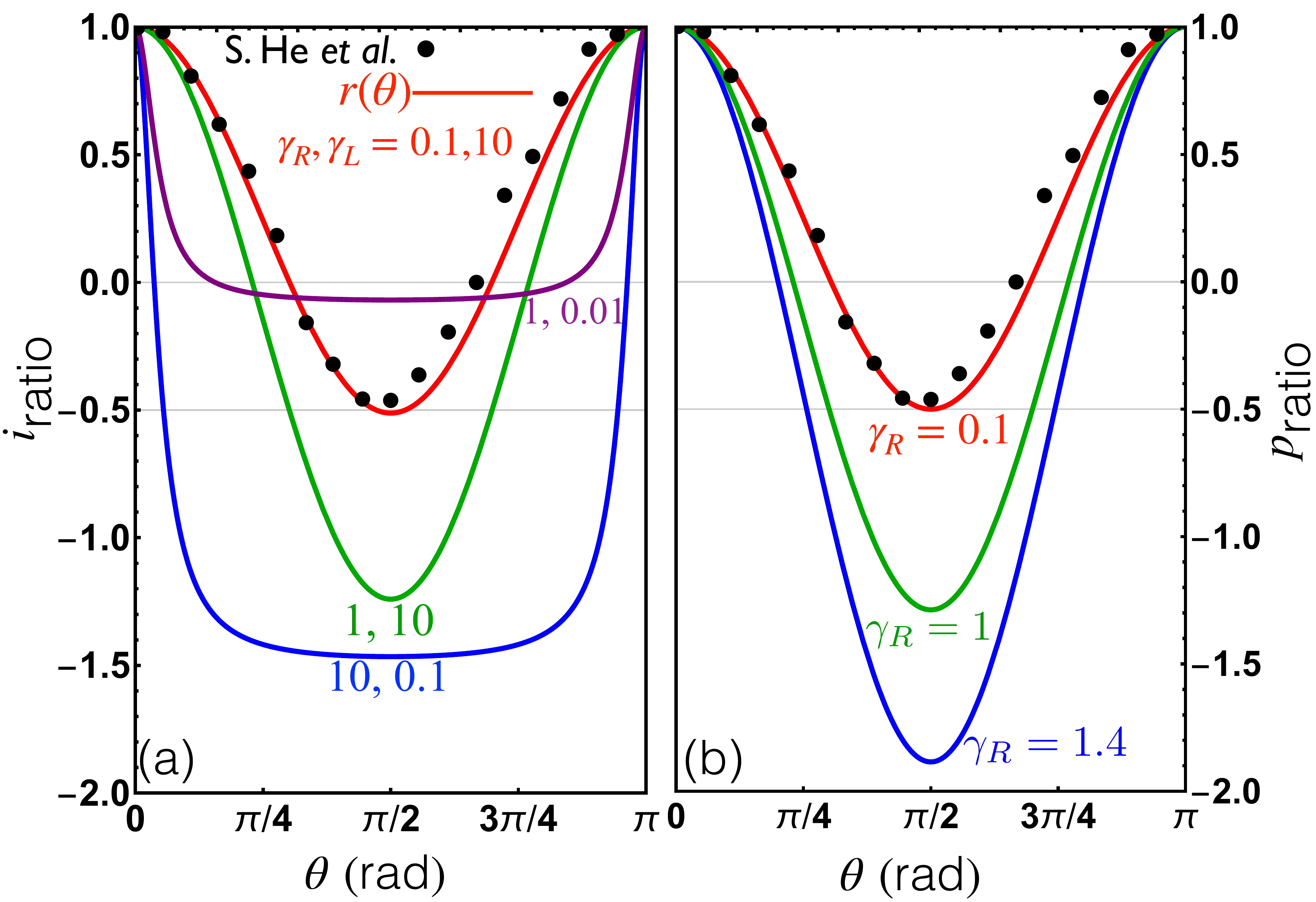}
        \caption[]
{
(a) $i_{\text{ratio}}$ and (b) $p_{\text{ratio}}$ as a function of applied field angle for a series of hopping rates (labeling each solid curve) between the spin center level and a left NM contact and right FM contact. Black circles are data from Ref.~\onlinecite{He2016}.  (a) Solid curves (red, purple, green, and blue) are determined from numerically averaging over the gaussian distribution of hyperfine fields for $i(0)$. 
(b) The results are independent of the left electrode $\rightarrow$ barrier trap hopping rate.  Solid curves (black, green, and blue) are determined from {Eq. (\ref{eq:pratio})}. The red numerical line for $\gamma_R = 0.1$, $\gamma_L = 10$ is nearly identical to the analytic function $r(\theta) = \frac{3}{2}\cos^2\theta - \frac{1}{2}$. 
}\label{fig:pRatio}
        \end{centering}
\end{figure}

\emph{Oblique Fields.}--- Recently, the effect of oblique magnetic fields was measured to help distinguish  spin accumulation and magnetic-field-dependent transport though barrier traps\cite{He2016}.
A  strong field  applied parallel to the magnetization suppresses any stray fields and returns the spin accumulation  to its stray-field-free value, which for oblique fields  is proportional to $\cos^2\theta$ for direct tunneling.\cite{Meier1984}
{Ref. \onlinecite{He2016} compared  oblique measurements (solid symbols in Fig. \ref{fig:pRatio}) to both the direct tunneling model and the trap model (spin bottleneck sites); the result was inconclusive as both models adequately described the data.}

We propose alternate quantities to distinguish the two models: the ratio of the current or spin accumulation at zero and high fields at angle $\theta$. The barrier trap model predicts a universal response whereas the stray-field model depends on the details of the magnetic layer's roughness.
Specifically,
\begin{equation}
i_{\text{ratio}} =  \frac{\langle i(b_0\rightarrow \infty, \theta) \rangle - \langle i(b_0 = 0)\rangle }{\langle i(b_0\rightarrow \infty, \theta = 0^{\circ})\rangle - \langle i(b_0 = 0)\rangle}
\end{equation}
for spin extraction and 
\begin{equation}
p_{\text{ratio}} =\frac{\langle p_z(b_0\rightarrow \infty, \theta) \rangle - \langle p_z(b_0 = 0) \rangle}{\langle p_z(b_0\rightarrow \infty, \theta = 0^{\circ}) \rangle - \langle p_z(b_0 = 0) \rangle}
\end{equation}
for spin injection. 
Angular brackets denote  averaging over the gaussian distribution of hyperfine fields.

In general $i_{\text{ratio}}$ can only be computed numerically but $p_{\text{ratio}}$ is analytically found to be \cite{supp}
\begin{equation}\label{eq:pratio}
p_{\text{ratio}} =  \frac{3 \mathcal{C} ( \cos^2\theta - 1)}{2} +1
\end{equation}
with
\begin{equation}
\mathcal{C} = \left[1 - \frac{\gamma_R^2}{\overline{b}_{hf}^2}+\frac{\sqrt{2\pi}}{2}\frac{\gamma_R^3}{\overline{b}_{hf}^3}e^{{\gamma_R^2}/{2\overline{b}_{hf}^2}} \text{erfc}(\frac{1}{\sqrt{2}}\frac{\gamma_R}{\overline{b}_{hf}})\right]^{-1}
\end{equation}
where, remarkably, there is no dependence on $\gamma_L$ nor $\tau_s$,  and no approximations have been made on the relative size of the various rates.
As shown in the Supplement\cite{supp},
for either metric, if $\gamma_R \ll \overline{b}_{hf} \ll  \gamma_L$ then $i_{\text{ratio}} = p_{\text{ratio}} \equiv r(\theta) = (3\cos^2\theta - {1})/{2}$.
Figure \ref{fig:pRatio} shows the oblique field data from Ref. \onlinecite{He2016} as well as $r(\theta)$ and the agreement with the barrier trap prediction is remarkable. The fixed ratio between currents at $90^{\circ}$ and $0^{\circ}$ [$r(90^{\circ})  = -1/2$] was already noted in STO/LAO/Co structures\cite{Inoue2015}.
Ratios near this value hold in some other experiments as well, but not all\cite{Jain2012, Txoperena2013, Tinkey2014}. 
The common occurrence of the ratio is a strong indicator that stray fields are often not the IHE mechanism.
Different structures and magnets possess varying degrees of roughness and thus varying stray field distribution which have no cause to agree with $r(90^{\circ})=-1/2$.

Whether the line shapes are determined by spin transport centers or spin bottleneck centers is difficult to ascertain based solely on the angle-dependent amplitudes.
The width for spin bottleneck centers, determined by the hyperfine coupling, is the same for parallel and perpendicular applied fields.\cite{supp}
This is not the case for spin transport centers when {$\tau_s \overline{b}_{hf} \gg 1$}; in this instance the width is governed by $\tau_s^{-1}$.
We expect that both impurity types occur (in addition to direct tunneling) so disentangling their contributions in measurements in the literature is not feasible within the scope of this article.
For $\gamma_R \ll \overline{b}_{hf} \ll \gamma_L$, both $i_{\text{ratio}}$ and $p_{\text{ratio}}$ approach $r$ but outside that strict constraint the two ratios are not identical [Fig.~\ref{fig:pRatio}(a) vs. Fig.~\ref{fig:pRatio}(b)]. 
$p_{\text{ratio}}$ is  more sensitive to $\gamma_R$ than $i_{\text{ratio}}$.

\begin{figure}[ptbh]
 \begin{centering}
        \includegraphics[scale = 0.3,trim = 0 0 0 0, angle = -0,clip]{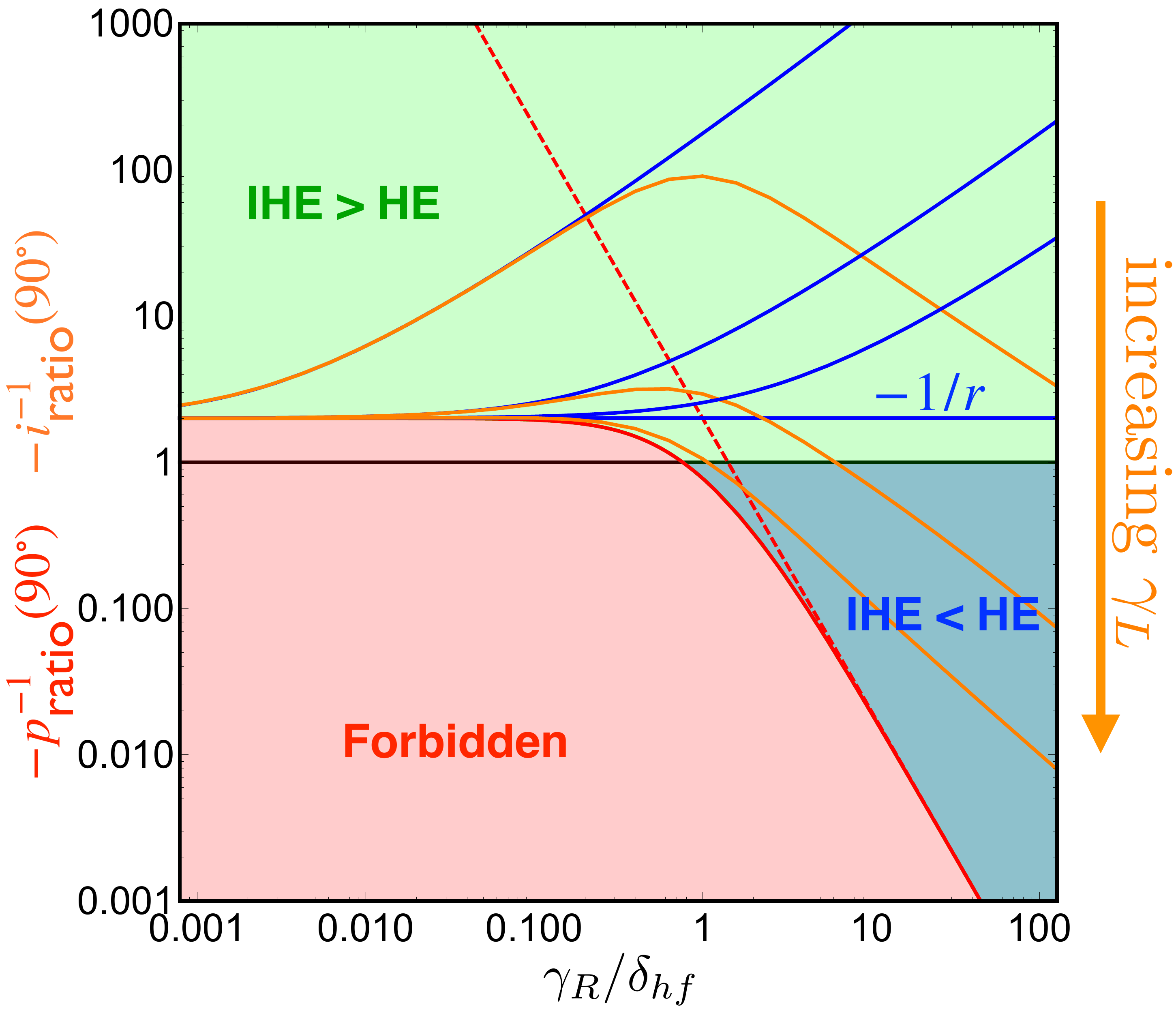}
        \caption[]
{
Ratio of the inverted Hanle effect (IHE) to the  Hanle effect (HE) absolute scale factors versus $\gamma_R$ for either spin extraction (orange, current ratio) or spin injection (red, polarization ratio). Blue curves are for $\gamma_R \ll \overline{b}_{hf}$  with the horizontal blue line being $-r^{-1}(\theta)$. The dashed red line is {$2\overline{b}_{hf}^2 \csc\theta/\gamma_R^2$}. 
Values of $\gamma_L$ are 0.001, 0.1, 1. As $\gamma_L$ increases further, the orange lines converges to the red line.  $p_{\text{ratio}}$  is independent of $\gamma_L$
}\label{fig:ratioPhaseDiagram}
        \end{centering}
\end{figure}

Figure \ref{fig:ratioPhaseDiagram} summarizes the magnitudes of the conventional and pseudo HE and IHE for either spin injection (red) or spin extraction (orange). The blue curves (red dashed line) indicate slow (fast) $\gamma_R$ approximations for extraction (injection). 
Three regions are represented by red (forbidden), cyan (IHE $<$ HE), and green (IHE $>$ HE). 
The quantity $-x^{-1}_{\text{ratio}}(90^{\circ})$ should be interpreted as {the ratio between IHE and HE: \emph{e.g.} $-i^{-1}_{\text{ratio}}(90^{\circ}) = \Delta i (\theta = 0^{\circ})/ \Delta i (\theta = 90^{\circ})  $}
The large-$\gamma_R$ asymptotic behavior of $-p^{-1}_{\text{ratio}}(\theta)$ is $2\overline{b}_{hf}^2 \csc\theta/\gamma_R^2$
%\begin{equation}\label{eq:asymptote}
%-1/p_{\text{ratio}} \approx 2\frac{\delta_{hf}^2}{\gamma_R^2}\csc\theta
%\end{equation}
which is the dashed red line in Figure \ref{fig:ratioPhaseDiagram}.

\emph{Ramifications for Non-Local Spin Detection.}---  For smaller voltage bias hops occur back and forth from each lead to the interfacial barrier trap. At zero bias (or zero charge current), no spin current is produced \cite{supp}. 
Thus the described magneto-effects for  spin transport centers or spin bottleneck %$s = 0$
centers cannot occur without a bias, so such effects are not expected in nonlocal measurements. Ref.~\onlinecite{Aoki2012} observes impurity-assisted signatures in three-terminal but not four-terminal devices. 
Thus the trap effects  here will not cause confusion for spin pumping and thermal spin transport experiments involving FM/NM interfaces without charge currents. The traps as described here will not alter the spin currents without charge currents, nor
 can unbiased spin centers mediate  ferromagnetic proximity polarization phenomenon\cite{Kawakami2001, Ciuti2002, Ou2016}.

\emph{Concluding Remarks.} ---
The ambiguity in the spin-dependent magnetoresistance in a resonant tunneling formulation originates from the dependence of the  Hubbard Hamiltonian ($Un_\uparrow n_\downarrow$) on the spin quantization axis\cite{Song2014,Yue2015b}. Our calculations and predictions are independent of the quantization axis and agree with Ref.~\onlinecite{Song2014}, although the reason why this is the ``proper'' choice  is unclear; we thus suggest that the SLE approach is more robust as no assumption of an axis is required.

\begin{acknowledgments}
The solution of these problems, comparison with experiment, and key results were supported by the U. S. Department of Energy, Office of Science, Office of Basic Energy Sciences, under Award Number DE-SC0016379.
The formulation of this problem and initial results were supported in part by C-SPIN, one of six centers of STARnet, a Semiconductor Research Corporation program, sponsored by MARCO and DARPA.
\end{acknowledgments}

%\bibliography{../../central-bibliography}
%\newpage

%\bibliography{central-bibliography}

%\include{supplement}

%%%%%% Everything commented out from here on out %%%%%%%%%%


\end{document}